\documentstyle[12pt,amssymb,epsfig]{article}
\begin{document}
\pagestyle{empty}
\renewcommand{\thefootnote}{\fnsymbol{footnote}}
\def\lsim{\raise0.3ex\hbox{$<$\kern-0.75em\raise-1.1ex\hbox{$\sim$}}}
\def\gsim{\raise0.3ex\hbox{$>$\kern-0.75em\raise-1.1ex\hbox{$\sim$}}}
\def\noi{\noindent}
\def\nn{\nonumber}
\def\bea{\begin{eqnarray}}  \def\eea{\end{eqnarray}}
\def\beq{\begin{equation}}   \def\eeq{\end{equation}}
\def\beeq{\begin{eqnarray}} \def\eeeq{\end{eqnarray}}
\def\R{ {\rm R \kern -.31cm I \kern .15cm}}
\def\C{ {\rm C \kern -.15cm \vrule width.5pt \kern .12cm}}
\def\Z{ {\rm Z \kern -.27cm \angle \kern .02cm}}
\def\N{ {\rm N \kern -.26cm \vrule width.4pt \kern .10cm}}
\def\1{{\rm 1\mskip-4.5mu l} }
\def\lsim{\raise0.3ex\hbox{$<$\kern-0.75em\raise-1.1ex\hbox{$\sim$}}}
\def\gsim{\raise0.3ex\hbox{$>$\kern-0.75em\raise-1.1ex\hbox{$\sim$}}}
\def\sq{\hbox {\rlap{$\sqcap$}$\sqcup$}}
\vbox to 2 truecm {}
\centerline{\Large \bf Hyperon Enhancement in the Dual Parton Model.}

\vskip 1 truecm
\centerline{\bf A. Capella and C. A. Salgado}
\centerline{Laboratoire de Physique Th\'eorique\footnote{Unit\'e Mixte de Recherche -
CNRS - UMR n$^{\circ}$ 8627}}  \centerline{Universit\'e de Paris XI, B\^atiment 210,
F-91405 Orsay Cedex, France}

\vskip 1 truecm 
\begin{abstract}

We review the two sources of hyperon enhancement in the dual parton model:
strings originating from diquark--antidiquark pairs in the nucleon sea and net
baryons containing two or three sea quarks with a yield controlled by the
observed stopping. We show that adding final state interactions (including
strangeness exchange reactions as well as the inverse reactions required by
detailed balance) with a single averaged cross--section $\sigma=0.2$ mb, we
can explain the observed hyperon enhancement in PbPb collisions at CERN SPS.

\end{abstract}

\vskip 1 truecm

\noindent LPT Orsay 00-66 \par
\noindent July 2000 \par

\newpage
\pagestyle{plain}
\baselineskip=24 pt

The dual parton model (DPM) is an independent string model. The total number of
strings is $2n$ where $n$ is the number of inelastic collisions. When a nucleon
undergoes a single inelastic scattering, the two produced strings have only
valence constituents (quark and diquark) at their ends. In the case of multiple
collisions the extra strings involve sea quarks or diquarks at their ends. We
have shown in \cite{cs2} and \cite{cfs1} that this produces a substantial
increase of the ratio of strange over non--strange baryons in $AA$ collisions
with increasing centrality. However, this mechanism produces equal enhancement
of baryons and antibaryons -- in disagreement with experiment. Fortunately,
there is another source of hyperon enhancement, which is intimately related to
baryon stopping, and enhances only the net hyperon yield $Y-\bar Y$. Indeed,
large number of net baryons are produced at mid-rapidities in central $AA$
collisions. It has been argued \cite{cs2}--\cite{capkop}
that they are dominantly
made out of the string junction (SJ) (see \cite{rossiven}), which carries the
baryon quantum number, plus three sea quarks (see Fig. 1). 
It is then obvious that a large number of net $\Lambda$, $\Xi$ and $\Omega$
(i.e. an increase of their yields per participant) will also take place. As a
matter of fact, this is the only possibility to produce net $\Omega$'s. The
experimental value $\bar\Omega/\Omega \sim 0.4$ \cite{wa97}
in central PbPb collisions at
mid--rapidities is very much in favour of the above picture. Moreover, there
will also be a substantial increase in the yield of $K^+$ associated to the
production of $\Lambda$'s.

\begin{figure}[tb]
\centerline{\epsfxsize=7cm\epsfbox{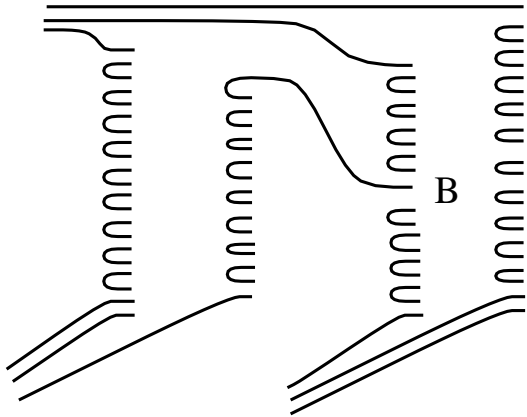}}
\caption[a] { \small Example of diquark breaking (DB) diagram for net baryon
production
in $pA$ with two inelastic collisions. The baryon is, in this case, made out of
three sea quarks.}
\end{figure}

The two sources of strangeness enhancement described above were studied in
detail in refs. \cite{cs2} and \cite{cfs1}. The numerical results are shown by
the dashed lines in Fig. 2. We see that the $p$ and $\Lambda$ yields are well
reproduced. The $\Xi$'s are slightly underestimated. However, the $\Omega$'s
are too low by a factor of 5.

\begin{figure}[tb]
\centerline{\epsfxsize=12cm\epsfbox{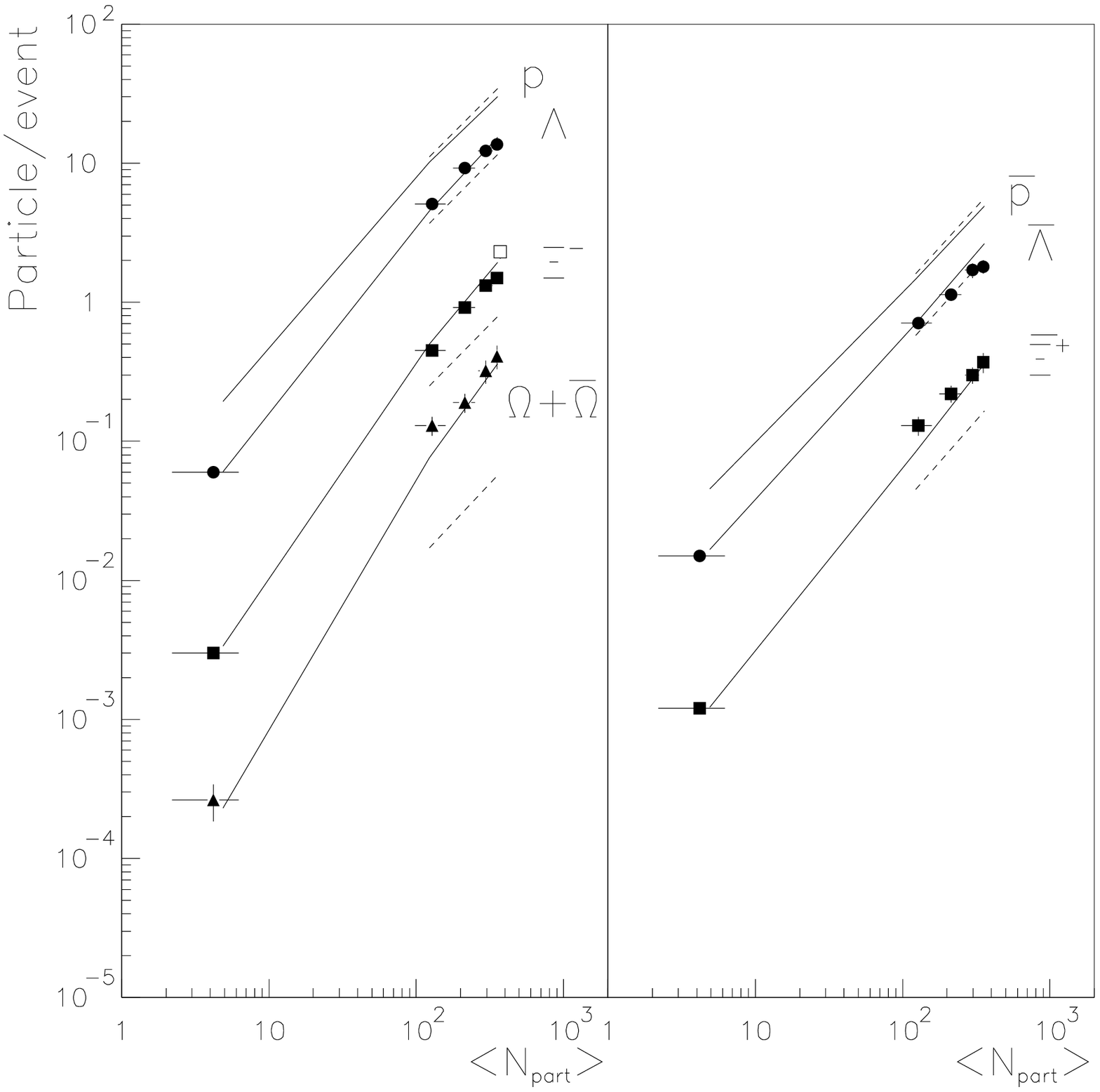}}
\caption[a] { \small Yields of $p$, $\Lambda$, $\Xi^-$, $\Omega+\bar\Omega$,
$\bar p$, $\bar\Lambda$ and $\bar\Xi^+$ for minimum bias $pPb$ (158 Gev/c)
and central $PbPb$ collisions (158 AGeV/c) in four centrality bins.
Experimental data are
 from WA97 \cite{wa97} (black points) and NA49 \cite{na49} (open
square).
Dashed lines are our results before final state interaction and
full lines are the results including this final state interaction.}

\end{figure}

In an attempt to describe the $\Omega$ yield, we have introduced in refs.
\cite{cs2} and \cite{cfs1}
the final state interactions~: 
\begin{equation}
\pi N \to K
\Lambda,\, \pi N \to K \Sigma,\, \pi \Lambda \to K \Xi,\, \pi \Sigma \to K
\Xi \, \,
\textnormal{and}\,\, \pi \Xi \to K \Omega, 
\label{eq1}
\end{equation}
\noindent
plus the corresponding reactions for the
antiparticles. They are governed by the gain and loss differential equations
\cite{16r}

\begin{equation}
\label{6e}
{dN_i \over d^4x} = \sum_{K, \ell} \sigma_{k\ell} \ \rho_k(x) \ \rho_{\ell}(x)
-
\sum_k \sigma_{ik} \ \rho_i(x) \ \rho_k(x) \ . \end{equation}

\noindent The first term in the r.h.s. of (\ref{6e}) describes the production
of particle $i$ resulting from the interaction of particles $k$ and $\ell$ with
space-time densities $\rho (x)$ and cross-sections $\sigma_{k\ell}$ (averaged
over the momentum distributions of the interacting particles). The second term
describes the loss of particle $i$ due to its interaction with particle $k$.
The initial densities are the ones obtained without final state interaction
and the averaged cross-sections are taken to be the same for all processes.
For details see \cite{cs2}.

Proceeding in this way, we were able to reproduce the observed enhancement of
all hyperon species \cite{cs2,cfs1}. However, in our approach we neglected the
inverse reactions, which are required by detailed balance, as well as the
charge exchange reactions:

\begin{equation}
\pi\Lambda \rightleftarrows KN,\ \pi\Sigma\rightleftarrows K\Lambda,\ 
\pi\Xi\rightleftarrows K\Lambda, \ \pi\Xi\rightleftarrows K\Sigma, \,\, 
\textnormal{and}\, \,\pi\Omega\rightleftarrows K\Xi.
\label{eq3}
\end{equation}
\noindent
together with the corresponding ones for antiparticles.

Although the possibility to neglect such reactions was qualitatively justified
using the relative size of the initial densities involved \cite{cs2,cfs1,capp},
this was considered to be a crucial drawback of our model \cite{heinz}. In view
of that, we have now introduced all these reactions. For simplicity we use a
single value for all averaged cross--sections $\sigma_{kl}=\sigma$. The results
with $\sigma=0.2$ mb are shown by the full lines of Fig. 2. We see that the
agreement with experiment is satisfactory. 

The reason why the introduction of the new reactions has produced no
substantial change in our former results is the following:
concerning the non strangeness enhancement reactions (inverse of those in
(\ref{eq1})) such as
$K\Lambda\rightarrow \pi N$, it is obvious that since $\rho_K<\rho_\pi$ and
$\rho_\Lambda<\rho_N$ one has $\rho_K\rho_\Lambda\ll \rho_\pi\rho_N$, and the
effect of these reactions is very small (of course, if the interaction
time would be much larger than the 6 fm measured from Bose-Einstein
interferometry, the inverse reactions would be crucial). For the strangeness
exchange reactions (\ref{eq3}) such as $\pi\Omega\to\bar K\Xi$, the situation
is different. If we compare this reaction with $\pi\Xi\to K\Omega$, the former
is disfavored since $\rho_\Xi>\rho_\Omega$. However, since $\Omega$'s are
strongly enhanced, the effect of the former might be important  -- and would
destroy $\Omega$'s. Actually, it turns out that the effect of this reactions is
only moderate. Moreover, the inverse reaction $\bar K\Xi\to \pi\Omega$ turns
out to be of comparable importance and restores the yield of $\Omega$'s
obtained without the strangeness--exchange reactions (\ref{eq3}).

In conclusion, we have shown that the dual parton model, supplemented with
final state interactions (both with and without strangeness exchange) describes
the observed enhancement of hyperon and antihyperon yields with a single value
of the averaged cross--sections -- which turns out to be rather small:
$\sigma=0.2$ mb.

\noi \subsection*{Acknowledgments}
\hspace{\parindent} It is a pleasure to thank N. Armesto,
J. A. Casado, E. G. Ferreiro, U. Heinz, A. B.
Kaidalov, C. Pajares
and J. Tran Thanh Van for discussions. 
A.C. acknowledges partial support from a NATO grant
OUTR.LG 971390. C.A.S. thanks Ministerio de Educaci\'on y Cultura of Spain for
financial support.

\end{document}